\documentclass[]{revtex4}
\usepackage{hyperref}
\usepackage{amsmath}
\usepackage[psamsfonts]{amssymb }
\usepackage{mathrsfs}
\begin{document}

\title{Aspects of finite field-dependent  symmetry in SU(2) Cho--Faddeev--Niemi decomposition }

\author{ Sudhaker Upadhyay}
 \email {sudhakerupadhyay@gmail.com; 
 sudhaker@boson.bose.res.in}

\affiliation { S. N. Bose National Centre for Basic Sciences,\\
Block JD, Sector III, Salt Lake, Kolkata -700098, India. }
 
\begin{abstract}
 In this Letter
 we consider $SU(2)$ Yang-Mills theory analysed in
 Cho--Faddeev--Niemi  variables
 which remains invariant under local gauge transformations.
 The BRST symmetries of this theory is generalized
 by making the infinitesimal parameter finite and field-dependent.
 Further, we show that under  appropriate choices of
 finite and field-dependent parameter,  the gauge-fixing and ghost terms corresponding to
 Landau as well as maximal Abelian gauge
 for such  Cho--Faddeev--Niemi  decomposed theory
 appear naturally within functional integral through
 Jacobian calculation.

  \end{abstract}
\maketitle

\section{  Introduction}
The original formulation
of Yang-Mills (YM) theory is suitable to explain the theory in the high-energy limit. In the high-energy limit,   YM theory is asymptotically
free and can be solved perturbatively.
However, in the low-energy region, it fails to describe the dynamics correctly due to strong coupling. In studying the low-energy dynamics of YM theory and quantum chromodynamics (QCD), it is important to extract the
most relevant (the dynamical  or topological)  degrees of freedom.

The Cho--Duan--Ge--Faddeev--Niemi decomposition, also called as Cho--Faddeev--Niemi (CFN) decomposition in literature,  of the non-Abelian
connection (gauge potential) of YM theory, was originally proposed  by Cho \cite{cho}, Duan and Ge \cite{du}, and Faddeev and Niemi \cite{faddeev} independently. 
CFN decomposition enables us to explain and understand some of the low-energy phenomena by separating the   topological defects in a gauge-invariant manner.
This has been done by introducing  a color vector field
to extract explicitly the magnetic
monopole as a topological degree of freedom from the gauge potential without introducing the fundamental scalar field in YM theory.

Quark  (color)  confinement
is a phenomenon in the low-energy dynamics due to strong interactions.
Quark confinement is believed to be explained by topological defects including magnetic monopoles,
vortices and merons. The monopole condensation   provides indispensable description 
  of the quark confinement through a dual Meissner effect  \cite{nam}. 
 It is conjectured that the restricted
part of QCD which comes from the ``Abelian projection"
of the theory to its maximal Abelian subgroup is
responsible for the dynamics of the dual Meissner effect \cite{cho}. The basic idea behind Abelian projection  is that the partial gauge-fixing can extract the physical degrees of freedom relevant to the long distance structure of QCD  \cite{hoof}. A magnetic monopole appears as a defect (singularity) of the partial gauge-fixing at degenerate points of the operator to be diagonalized through the Abelian projection. The most efficient partial gauge-fixing from this  point of view
is known to be the maximally Abelian gauge (MAG) \cite{kron}, although  there are
many candidates for Abelian gauge  \cite{van}. The numerical simulations \cite{suz} have confirmed that only MAG leads to the Abelian dominance   in the theory  at a long-distance scale \cite{eza} and also the magnetic monopole dominance for the string tension \cite{sta} .

The BRST transformation plays a very important role in the proof of  renormalizability  and unitarity  of the gauge
theories \cite{brst,tyu,ht,wei}. The BRST quantization of the CFN decomposed YM theory
in the continuum formulation has been studied thoroughly \cite{kon}.
To determine the nilpotent
BRST transformations for all the fields the nilpotency property
has been utilized \cite{kon}. At this place, one needs
 more ghost and antighost fields  than the usual YM theory 
reflecting the enlarged local gauge symmetry of the CFN decomposed
YM theory. It has also been discussed that in order to remove the redundancy
of the gauge degrees of freedom
of this theory, it is necessary to incorporate the Landau covariant gauge condition also in addition to the new MAG.
 
 The generalization of BRST transformation
 by making the infinitesimal parameter finite and field-dependent is known as
 finite field-dependent BRST (FFBRST) transformation \cite{bm}.
Such generalizations have found various applications in gauge field theories 
 \cite{bm,sb,sb2,sdj1,rb,susk,jog,sb1,smm,fs,sud1,rbs}.
 For example, a correct prescription for   poles in the gauge field propagators in noncovariant gauges has been derived by connecting  covariant gauges and noncovariant gauges of the theory by using FFBRST transformation \cite{jog}.
The Gribov-Zwanziger theory \cite{gri,zwan}  which
is free from Gribov copies  plays a crucial role in the non-perturbative
low-energy region while it can be neglected in the  perturbative high-energy
region, has also been related to the YM theory  in Euclidean space through FFBRST transformation \cite{sb, sb2}. However, such formulation has not been 
investigated for YM theory explaining low-energy dynamics. With this motivation
we implement this formulation
in the CFN decomposed $SU(2)$ YM theory.
 
 In this Letter, we generalize the BRST transformations for CFN decomposed YM theory. This
 is achieved, first by making all the fields $\kappa$ (a parameter) dependent
 through continuous interpolation. Then, the infinitesimal 
 constant parameter characterizing the BRST symmetry is made
 field-dependent. After that, we integrate such
 infinitesimal field-dependent parameter and therefore  the finite field-dependent 
 BRST parameter is obtained. Further, we construct 
 the FFBRST transformations for CFN decomposed YM theory characterized 
 by an arbitrary finite field-dependent parameter.
Such FFBRST transformations are symmetry of the effective action
 only but not of the partition function
 because the path integral measure is not
 covariant under these. The Jacobian 
 of the path integral measure in the expression of partition function changes non-trivially
 under such FFBRST transformation. 
 We show that for an appropriate choice of
 this parameter the Jacobian of path integral measure leads to
 gauge-fixing and corresponding ghost terms in the
 theory naturally. Therefore we claim that the
 gauge-fixing and ghost terms are contribution of Jacobian of path integral measure 
 under FFBRST transformation.  
 
The Letter is presented as follows. In Sec. II, we study the preliminaries
about CFN decomposition of YM theory. The nilpotent BRST symmetry
is also discussed in this section. Sec. III is devoted to study the
methodology used in generalizing BRST symmetry under which Jacobian changes
non-trivially. The discussion about the evaluation of the non-trivial Jacobian
is made in section 
IV. Further, in Sec. V, we show the emergence of gauge-fixing and ghost terms
of CFN decomposed theory naturally. Last section  is reserved for
discussions and conclusions.

\section{The CFN decomposition of Yang-Mills theory}
In this section, we discuss the Cho--Faddeev--Niemi (CFN) decomposition of 
$SU(2)$ YM theory, which explains the infra-red limit of the theory,
and its BRST invariance. To achieve the CFN decomposition of connection
$A_\mu(x)$ of YM theory,   the connection is separated in Abelian projection
(called as restricted potential) part and in remaining gauge covariant potential part
because Abelian projection dominates in infra-red limit.
The Abelian
projection of connection is achieved by introducing a three
component vector $\hat n(x)$ of unit length, i.e.,
\begin{equation}
 \hat n(x)\cdot\hat n(x)  = 1
  \label{nn}
\end{equation}
  Now, the CFN decomposition of 
  the connection $A_\mu$  in terms of component vector $\hat n$
is given by,
\begin{equation}
A_\mu(x)
 =\hat A_\mu(x)  \hat n(x)
  + g^{-1}\partial_\mu \hat n(x)\times \hat n(x)
  + X_\mu(x),\label{a}
\end{equation}
where $g$ is coupling constant, $\hat A_\mu(x)$ is the Abelian component
(the
electric potential)
of the connection  and 
$ X_\mu(x)$ is gauge covariant potential  orthogonal to $\hat n(x)$,
i.e.,
\begin{equation}
\hat n(x) \cdot  X_\mu(x) = 0.\label{or}
\end{equation}
  The connection $A_\mu$  given in Eq. (\ref{a}) is further
  written in terms of Abelian projected 
  vector field $V_\mu$ for later convenience  as
\begin{equation}
A_\mu(x)=  V_\mu(x) +  X_\mu(x),
\end{equation}
where  the  Abelian projection  $ V_\mu(x)$ is defined by
\begin{equation}
 V_\mu(x)  
    = \hat A_\mu(x) \hat n(x)
  + g^{-1}\partial_\mu  \hat n(x)\times  \hat n(x).
\end{equation}
This Abelian projection (restricted potential) leaves the component vector $\hat n$ 
invariant under parallel transport
\begin{equation}
D[ V]\hat n:=\partial_\mu \hat n+g A_\mu(x) \times\hat n =0.\label{par}
\end{equation} 
Now, exploiting equations (\ref{a}) and (\ref{or}),  the
electric potential $\hat A_\mu(x)$ is expressed 
in terms of $\hat n$ and $A_\mu$ as follows:
\begin{equation}
\hat A_\mu(x) ={  \hat n}(x)\cdot A_\mu(x).\label{c}
\end{equation}
Further,   relation (\ref{par}) reflects that the  gauge covariant potential 
$X_\mu$ also depends only on $\hat n$ and $A_\mu$ as \cite{ss}
\begin{equation}
 X_\mu(x)
  =g^{-1}\hat n(x)\times D_\mu[A]\hat n(x).
\label{x}
\end{equation}
Incorporating these CFN variables, the classical Lagrangian density for YM theory   is defined by
\begin{equation}
{\cal L}_{YM} =-\frac{1}{4} \hat{F}_{\mu\nu}^2, \label{ym}
\end{equation}
where the field-strength tensor $\hat{F}_{\mu\nu}$ has the following expression:
\begin{eqnarray}
\hat{F}_{\mu\nu}= \left[ 
\partial_\mu \hat A_\nu -\partial_\nu \hat A_\mu -g^{-1} \hat n \cdot (\partial_\mu \hat n \times \partial_\nu \hat n)\right]\hat n + D_\mu[V] X_\nu -
D_\nu[V] X_\mu +g  X_\mu \times  X_\nu.
\end{eqnarray}
The above Lagrangian density described in terms of CFN variables  remains invariant  under following
 local gauge symmetry:
\begin{eqnarray}
\delta A_\mu(x)
  &=&  D_\mu[A]{\omega}(x) ,\nonumber\\
\delta{\hat n}(x)
  &= & g{ \hat n}(x) \times { \theta}(x) 
  =g{\hat n}(x) \times { \theta}_\perp(x) ,\label{gauge}
\end{eqnarray}
where ${\omega}(x)$ is an arbitrary local parameter and
 $\theta(x)$ is an angle of local rotation with component ${\theta}_\perp(x)$
which is orthogonal to $\hat n$, i.e., $\hat n\cdot {\theta}_\perp(x) =0$.
Now, using equations (\ref{c}), (\ref{x}) and (\ref{gauge}),
we are able to write the gauge transformations of   variables 
$\hat A_\mu $ and $ X_\mu  $ as follows:
\begin{eqnarray}
  \delta \hat A_\mu(x) 
  &=&   g( \hat n(x) \times A_\mu(x)) \cdot ({\omega}_\perp(x) - {\theta}_\perp(x)) + \hat n(x) \cdot \partial_\mu {\omega}(x) ,\nonumber
\\
  \delta  X_\mu(x) 
  &=&  g  X_\mu(x) \times  ({\omega}_\parallel(x)+{\theta}_\perp(x)) + D_\mu[ V]({\omega}_\perp(x)-{\theta}_\perp(x))  ,
  \label{gt}
\end{eqnarray}
where ${}_\parallel$ and ${}_\perp$ denote the components of variables parallel and perpendicular to $\bm{n}$. 
In this context, two types of  local gauge transformations, the active
(background)
gauge transformation
and the passive (quantum)
gauge transformation have been studied \cite{bae}.
The gauge transformations in Eqs. (\ref{gauge}) and (\ref{gt}) are identified
as the active gauge transformation. 
For particular case: $\bm\omega_\perp(x)=\bm\theta_\perp(x)$,
these active transformations given (\ref{gauge}) and (\ref{gt}) reduce to  
\begin{eqnarray}
  \delta_{\omega'}   \hat n(x)  &=& g  \hat n(x) \times  {\omega'}(x)  ,
\nonumber\\
 \delta_{\omega'} \hat A_\mu(x) &=&     \hat n (x)\cdot \partial_\mu  {\omega'} (x)  ,
\nonumber\\
  \delta_{\omega'}  X_\mu (x)&=&  g  X_\mu (x)\times  {\omega'} (x),
\nonumber\\
\delta_{\omega'}  V_\mu(x) &=&   D_\mu[ V] {\omega'} (x),\label{g1}
\end{eqnarray}
with the local parameter $ \omega'(x)=( \omega_\parallel(x), \omega_\perp (x)= \theta_\perp(x))$. 
However,  
the passive gauge transformation are defined by  \cite{bae}
\begin{eqnarray}
  \delta_\omega  \hat n & =& 0  ,\nonumber
\\
 \delta_\omega \hat A_\mu &=& \hat n \cdot  D_\mu[A]  {\omega} ,\nonumber
\\
  \delta_\omega  X_\mu &=&     D_\mu[A]  {\omega} -  \hat n( \hat n \cdot  D_\mu[A] {\omega}) ,\nonumber
\\ 
     \delta_\omega  V_\mu 
&= &  \hat n( \hat n \cdot  D_\mu[A] {\omega})  . \label{g2}
\end{eqnarray}
Here we note that the above two gauge transformations given in equations (\ref{g1}) and (\ref{g2}) are not independent. 

\subsection{Gauge-fixed action and BRST symmetry}
The presence of gauge symmetries in the YM theory decomposed 
in CFN variables reflects 
that theory is enriched with some redundant degrees of freedom. It is also
 obvious 
from LHS and RHS of relation (\ref{a}) that
RHS has two extra degrees of freedom introduced by 
$\hat n$. To fix these two redundant degrees of freedom  we put an extra constraints on $X_\mu$ as
\begin{equation}
D_\mu [V]X^\mu =0,
\end{equation}
which is called as new MAG \cite{kon, kon1}.
The nilpotent BRST symmetry for YM theory  decomposed in CFN variables
has been discussed in great details  \cite{kon}.
To write the BRST symmetry transformations, we introduce two
ghost fields ${ \mathbb C}_\omega$ and ${ \mathbb C}_\theta$ corresponding to 
the parameters
$\omega$ and $\theta$, respectively, characterising gauge transformations. 
Then, the BRST transformations for
$A_\mu$ and $\hat n$,  by replacing   $\omega$ and $\theta$ 
with ${\mathbb C}_\omega$ and ${\mathbb C}_\theta$  respectively  in (\ref{gauge}),
is defined as
  \begin{eqnarray}
  s_b A_\mu =- D_\mu [A] { \mathbb C}_\omega,\ \
s_b \hat n =- g\hat n \times   {\mathbb C}_\theta =-g\hat n \times   {\mathbb C}_\theta^\perp,\label{brs0}
  \end{eqnarray}
  where we have acquired the fact that
  $\hat n\cdot {\mathbb C}_\theta=0$.
  
\underline{BRST symmetry of $\omega$ sector}:

  To analyse the BRST transformation for fields in $\omega$ sector
  we use mainly the nilpotency property of BRST operator.
    The nilpotency of the BRST symmetry for $A_\mu$ reflects the
  BRST of ghost field $ {\mathbb C}_\omega$ as follows \cite{kon}:
  \begin{equation}
  s_b  {\mathbb C}_\omega =-\frac{g}{2} {\mathbb C}_\omega \times 
  {\mathbb C}_\omega,\label{brs1}
  \end{equation}
One can easily check the nilpotency of $ {\mathbb C}_\omega $, i.e.,
$  s_b^2  {\mathbb C}_\omega =0$.
Making analogy to standard YM case, we write
the BRST symmetry for the antighost field $\bar {\mathbb C}_\omega$ by
 \begin{eqnarray}
 s_b \bar {\mathbb C}_\omega = i B_\omega,\ \ s_b B_\omega =0,\label{brs2}
 \end{eqnarray}
where $B_\omega$ is Nakanishi-Lautrup type auxiliary field. The BRST
 transformation of $B_\omega$ is an outcome of  nilpotency
property of BRST transformation for $\bar {\mathbb C}_\omega$.

\underline{BRST symmetry of $\theta$ sector}:

The BRST symmetry of the fields in $\theta$ sector can be written as 
follows:
\begin{eqnarray}
s_b {\mathbb C}_\theta &=&g{\mathbb C}_\theta \times{\mathbb C}_\theta,\nonumber\\
s_b \bar{\mathbb C}_\theta &=& iB_\theta,\nonumber\\
s_b B_\theta &=& =0,\label{brs3}
\end{eqnarray}
 where $B_\theta $ is Nakanishi-Lautrup type auxiliary field for
 $\theta$ sector.
 
With the help of  BRST transformations given in  Eqs. (\ref{brs0}), (\ref{brs1})
(\ref{brs2}) and (\ref{brs3}), we are able to
to write the BRST transformations for CFN variables $\hat A_\mu $ 
and $ \mathbb{X}_\mu $ 
\begin{eqnarray}
 s_b \hat A_\mu  
 & =&  -g(\hat{n} \times  {A}_\mu ) \cdot (\mathbb C_\omega^\perp - \mathbb C_\theta^\perp ) - \hat {n} \cdot \partial_\mu \mathbb C_\omega ,\nonumber
\\
  s_b {X}_\mu 
 & =& -g {X}_\mu \times  (\mathbb C_\omega^\parallel +\mathbb C_\theta^\perp ) - D_\mu[ {V}](\mathbb C_\omega^\perp -\mathbb C_\theta^\perp ),\label{brs4}
 \end{eqnarray}
 where ${}_\parallel$ and ${}_\perp$ denote the components which are  parallel and perpendicular to
 $\hat {n}$ respectively.
 Now, the gauge-fixing and ghost parts of the  YM Lagrangian density
 described in CFN variables is given by \cite{kon}
 \begin{eqnarray}
 {\cal L}_{  GF+FP}
 ={\cal L}_{  GF+FP}^\omega  +{\cal L}_{  GF+FP}^\theta,\label{gg}
 \end{eqnarray}
 where the gauge-fixing and ghost terms for $\omega$ and $\theta$
 sectors are defined, respectively, as
 \begin{eqnarray}
 {\cal L}_{GF+FP}^\omega &=&  B_\omega \cdot \partial^\mu A_\mu  +i \bar{ \mathbb 
C}_\omega \partial^\mu D_\mu [A]{ \mathbb C}_\omega,\nonumber\\
{\cal L}_{GF+FP}^\theta &=& B_\theta \cdot D^\mu[V]  X_\mu -i  
\bar{ \mathbb C}_\theta \cdot D^\mu[ V-  X]D_\mu[ V+ X]  ({ \mathbb C}_\theta -{ 
\mathbb C}_\omega ).
 \end{eqnarray}
$ {\cal L}_{GF+FP}$ remains invariant under above
mentioned sets of BRST transformations. 
 Now, we are able to define the 
 partition function (vacuum to vacuum transition amplitude) 
 for CFN decomposed YM theory  in
 Euclidean space as follows:
 \begin{equation}
 Z[0] =\int D\phi\ e^{-S_{eff}},\label{zfun}
 \end{equation}
 where $D\phi$ is path integral measure written compactly
 in terms of  generic field $\phi (\equiv \hat n, \hat A_\mu,   
\mathbb{C}_\omega, \bar{\mathbb{C}}_\omega, B_\omega,
\mathbb{C}_\theta, \bar{\mathbb{C}}_\theta, B_\theta,
X_\mu)$. The  
 effective action for 
 YM theory decomposed in CFN variables, $S_{eff}$, is defined as a sum of
 classical part, gauge-fixing part and ghost part, i.e.,
\begin{equation}
S_{eff} =\int d^4 x\left[{\cal L}_{YM}+ {\cal L}_{GF+FP}\right].\label{action}
\end{equation} 
These partition function and effective action  are invariant under BRST transformations given in
equations (\ref{brs0}), (\ref{brs1}), (\ref{brs2}), (\ref{brs3}) and (\ref{brs4}).
\section{ Generalized BRST symmetry in Euclidean space}
In this section, we generalize the sets of BRST transformations (obtained in previous 
section) 
written for CFN variables by making the
infinitesimal Grassmann parameter finite and field-dependent.
For this purpose, we first define the infinitesimal BRST transformations ($\delta_b$)
written, compactly, for generic field $\phi (\equiv \hat n, \hat A_\mu,   
\mathbb{C}_\omega, \bar{\mathbb{C}}_\omega, B_\omega,
\mathbb{C}_\theta, \bar{\mathbb{C}}_\theta, B_\theta,
X_\mu)$, 
\begin{equation}
\delta_{b}\phi= s_b \phi\ \delta\Lambda,
\end{equation}
where $\delta\Lambda$ is an infinitesimal, space-time independent
parameter which belongs to Fermi-Dirac statistics. Now,
we make $\delta\Lambda$ finite and field-dependent without affecting its properties.
Such a generalization of BRST
transformation is known as finite field-dependent BRST (FFBRST) transformation \cite{bm}.
The mechanisms of FFBRST transformation are as follows: first of all
we start by making the infinitesimal
parameter  ($\delta\Lambda$)  field-dependent with introduction of an arbitrary parameter $\kappa (0\leq \kappa \leq 1)$, i.e., $\delta\Lambda[\phi(x,\kappa)]$, where all the fields generically $\phi (x)$ of the theory depend on $\kappa$ such that
$\phi (x, \kappa =0) =\phi (x)$ are initial fields and
$\phi (x, \kappa =1) =\phi' (x)$ are FFBRST transformed fields.

Now, the infinitesimal   field-dependent BRST transformations for
generic fields $\phi$ are defined explicitly as \cite{bm},
\begin{eqnarray}
\frac{d\phi(x, \kappa)}{d\kappa} = 
s_b \phi (x) \Lambda'[\phi(x,\kappa)], \label{infi}
\end{eqnarray}
with  infinitesimal field-dependent parameter $\Lambda'[\phi(x,\kappa)] =d(\delta\Lambda[\phi(x,\kappa)])/d\kappa$.
The
FFBRST transformation ($\delta_{f}$) is then  
constructed by integrating the above infinitesimal
field-dependent transformation from $\kappa = 0$ to 
$\kappa = 1$, i.e.,
\begin{eqnarray}
\delta_f \phi(x):= \phi'(x, \kappa =1) -\phi (x, \kappa=0) =s_b \phi(x) \Lambda
[\phi(x)],
\end{eqnarray}
where 
\begin{equation}
\Lambda[\phi(x)] = \int_0^1 d\kappa\ \Lambda' [\phi(x,\kappa)],\label{para}
\end{equation}
 is finite field-dependent parameter. 
 
 Following this methodology,  we construct the FFBRST transformation
 for CFN decomposed YM theory as follows:
 \begin{eqnarray}
\delta_f A_\mu &=&- D_\mu [A] { \mathbb C}_\omega\ \Lambda[\phi(x)],\nonumber\\
\delta_f\hat n &=&- g\hat n \times   {\mathbb C}_\theta\ \Lambda[\phi(x)]=-g\hat n \times   {\mathbb C}_\theta^\perp\ \Lambda[\phi(x)],\nonumber\\
 \delta_f  {\mathbb C}_\omega &=&-\frac{g}{2} {\mathbb C}_\omega \times 
  {\mathbb C}_\omega\ \Lambda[\phi(x)],\nonumber\\
 \delta_f \bar {\mathbb C}_\omega &=& i B_\omega\ \Lambda[\phi(x)],\ \ \delta_f B_\omega =0,\nonumber\\
\delta_f {\mathbb C}_\theta &=&g{\mathbb C}_\theta \times{\mathbb C}_\theta\ \Lambda[\phi(x)],\nonumber\\
\delta_f \bar{\mathbb C}_\theta &=& iB_\theta\ \Lambda[\phi(x)], \ \ 
\delta_f B_\theta = =0,\nonumber\\
\delta_f \hat A_\mu  
 & =&  -\left(g(\hat{n} \times  {A}_\mu ) \cdot (\mathbb C_\omega^\perp - \mathbb C_\theta^\perp ) + \hat {n} \cdot \partial_\mu \mathbb C_\omega \right)\ \Lambda[\phi(x)],\nonumber
\\
\delta_f {X}_\mu 
 & =& -\left(g {X}_\mu \times  (\mathbb C_\omega^\parallel +\mathbb C_\theta^\perp ) + D_\mu[ {V}](\mathbb C_\omega^\perp -\mathbb C_\theta^\perp )\right)\ \Lambda[\phi(x)],  \label{ffbrs}
 \end{eqnarray}
 where $\Lambda[\phi(x)]$ is an arbitrary finite field-dependent
 parameter. 
The  effective action given in Eq. (\ref{action}) is invariant 
under this FFBRST transformations, however, the
generating functional given in (\ref{zfun}) is not 
because the Jacobian of path integral measure in the expression of
partition function gives some non-trivial
contribution to it (for detail see e.g. \cite{bm}).

\section{Method for Evaluating the  Jacobian} 
For the symmetry of the generating functional we need to calculate the Jacobian of the 
path integral measure in the definition of generating functional.
The Jacobian  of the path integral measure for FFBRST transformation $J $  can be evaluated for some 
particular choices of the finite field dependent parameters $\Lambda[\phi(x)]$. We start with the definition of path integral measure, \cite{bm} 
\begin{eqnarray}
D\phi  &=&J( \kappa)\ D\phi(\kappa) \nonumber\\ 
&=& J(\kappa+d\kappa)\ D\phi(\kappa+d\kappa).
\end{eqnarray} 
 Now the transformation from $\phi(\kappa)$ to $\phi(\kappa +d\kappa)$ is infinitesimal 
in nature, thus the infinitesimal change in Jacobian can be calculated as  \cite{bm} 
\begin{equation}
\frac{J (\kappa)}{J (\kappa +d\kappa)}=\Sigma_\phi\pm 
\frac{\delta\phi(x, \kappa)}{
\delta\phi(x, \kappa+d\kappa)}, 
\end{equation}
where $\Sigma_\phi $ sums over all fields involved in the path integral measure  
and  $\pm$ 
sign refers to whether $\phi$ is a bosonic or a fermionic field.
Using the Taylor expansion we calculate the above expression as \cite{bm} 
\begin{eqnarray} 
 \frac{1}{J (\kappa)}\frac{dJ (\kappa)}{d\kappa} 
&=& - \int d^4x\left [\Sigma_\phi (\pm) s_b  
\phi (x,\kappa )\frac{
\partial\Lambda^\prime [\phi (x,\kappa )]}{\partial\phi (x,\kappa )} \right  ].\label{jac}
\end{eqnarray}
The Jacobian, $J(\kappa )$, can be replaced (within the functional integral)
in Euclidean space as \cite{bm, sb}
\begin{equation}
J(\kappa )\rightarrow e^{- (S_1[\phi(x,\kappa), \kappa] )},\label{js}
\end{equation}
if and only if the following condition is satisfied \cite{sb}
\begin{eqnarray}
\int  D\phi (x)  \left [ \frac{1}{J (\kappa )}\frac{dJ (\kappa )}{d\kappa} 
+\frac
{d  S_1[\phi (x,\kappa )]}{d\kappa} \right ]   e^{ -(S_{eff} +S_1) }=0, 
\label{mcond}
\end{eqnarray} 
where $ S_1[\phi ]$ is some local functional  of fields and satisfies the initial
condition
 \begin{equation}
S_1[\phi(\kappa =0) ]=0.\label{inco}
\end{equation}
\section{Emergence of gauge-fixing and ghost terms in the theory}  
 
In this section, we calculate the Jacobian of path integral measure
 under FFBRST transformation
given in Eq. (\ref{ffbrs}) for a particular choice of finite field-dependent parameter. With this particular choice of finite parameter 
the  gauge-fixing and Faddeev-Popov ghost terms emerge naturally
trough Jacobian calculation
within functional integration  
written in $SU(2)$ CFN variables. For this purpose, we make an ansatz for
finite field-dependent parameter obtainable from
following infinitesimal field-dependent parameter
\begin{eqnarray}
\Lambda '[\phi(y, \kappa)] & =&i\int d^4 y \left[\bar{ \mathbb C}_\omega\cdot \partial^\mu A_\mu
+\bar{ \mathbb C}_\theta\cdot  D^\mu[V]  X_\mu
\right],\label{thet}
\end{eqnarray}
using relation (\ref{para}). Now, exploiting relation (\ref{jac}) 
we calculate an infinitesimal change in Jacobian with respect to $\kappa$ 
for above $\Lambda '[\phi(y, \kappa)] $ as follows:
\begin{eqnarray}
\frac{1}{J(\kappa)}\frac{dJ(\kappa)}{d\kappa}&=&\int d^4 x \left [-  B_\omega \cdot \partial^\mu A_\mu -i 
\bar{ \mathbb C}_\omega \partial^\mu D_\mu [A]{ \mathbb C}_\omega -   B_\theta \cdot D^\mu[V]  X_\mu \right.\nonumber\\
&+ &\left. i
\bar{ \mathbb C}_\theta \cdot D^\mu[ V-  X]D_\mu[ V+ X]  ({ \mathbb C}_\theta -{ 
\mathbb C}_\omega )
\right].\label{jaco}
\end{eqnarray}
The expression of arbitrary functional $S_1[\phi(x,\kappa), \kappa]$,
which appears in the exponential of Jacobian given in Eq. (\ref{js}), is constructed as
\begin{eqnarray}
S_1 [\phi(x,\kappa), \kappa] &=&\int d^4 x \left [\chi_1(\kappa) B_\omega \cdot \partial^\mu A_\mu  +\chi_2 (\kappa) \bar{ \mathbb 
C}_\omega \partial^\mu D_\mu [A]{ \mathbb C}_\omega +
\chi_3(\kappa) B_\theta \cdot D^\mu[V]  X_\mu  \right.\nonumber\\
&+ &\left. \chi_4 (\kappa)   
\bar{ \mathbb C}_\theta \cdot D^\mu[ V-  X]D_\mu[ V+ X]  ({ \mathbb C}_\theta -{ 
\mathbb C}_\omega )
\right],\label{s1}
\end{eqnarray}
where $\chi_i (i=1,2,3,4)$ are arbitrary $\kappa$-dependent constants
and satisfy the boundary conditions
\begin{equation}
\chi_i (\kappa =0)=0.\label{ka}
\end{equation}
To calculate the exact values of these constants, we first calculate
the small change in above $S_1$ with respect to
 parameter $\kappa$  
 using equation (\ref{infi}) as,
\begin{eqnarray}
\frac{dS_1[\phi(x,\kappa), \kappa] }{d\kappa} &=&\int d^4 x \left [\chi_1^\prime(\kappa) B_\omega \cdot \partial^\mu A_\mu  
+\chi_2^\prime (\kappa) \bar{ \mathbb 
C}_\omega \partial^\mu D_\mu [A]{ \mathbb C}_\omega  +\chi'_3(\kappa) B_\theta \cdot D^\mu[V]  X_\mu \right.\nonumber\\
&
+&\left.\chi'_4 (\kappa)   
\bar{ \mathbb C}_\theta \cdot D^\mu[ V-  X]D_\mu[ V+ X]  ({ \mathbb C}_\theta -{ 
\mathbb C}_\omega )- (\chi_1 +i\chi_2)  B_\omega
\partial^\mu D_\mu [A]{ \mathbb C}_\omega \Lambda ' \right.\nonumber\\
&
+& \left. (\chi_3 -i\chi_4)  N_\theta \cdot D^\mu[ V-  X]D_\mu[ V+ X]  ({ \mathbb C}_\theta -{ 
\mathbb C}_\omega )\Lambda ' \right],\label{diff}
\end{eqnarray}
where  prime denotes the differentiation with respect to $\kappa$.
Then we put the condition for existence of above functional $S_1$ as an exponent of Jacobian $J$, i.e.,  the Eqs. (\ref{jaco}) and (\ref{diff}) must satisfy 
the condition given in  (\ref{mcond}). Therefore, this
reflects
\begin{eqnarray}
&&\int d^4 x \left [(\chi_1' -1) B_\omega \cdot \partial^\mu A_\mu  
+  (\chi_2' -i) \bar{ \mathbb 
C}_\omega \partial^\mu D_\mu [A]{ \mathbb C}_\omega  + (\chi_3' -1) B_\theta \cdot D^\mu[V]  X_\mu \right.\nonumber\\
&&
+\left. (\chi_4' +i)   
\bar{ \mathbb C}_\theta \cdot D^\mu[ V-  X]D_\mu[ V+ X]  ({ \mathbb C}_\theta -{ 
\mathbb C}_\omega )- (\chi_1 +i\chi_2)  B_\omega
\partial^\mu D_\mu [A]{ \mathbb C}_\omega \Lambda ' \right.\nonumber\\
&&
+  \left. (\chi_3 -i\chi_4)  N_\theta \cdot D^\mu[ V-  X]D_\mu[ V+ X]  ({ \mathbb C}_\theta -{ 
\mathbb C}_\omega )\Lambda ' \right] =0.
\end{eqnarray}
The comparison of coefficients from the terms of LHS  
to RHS of the above equation gives the following restrictions on
the parameters ($\chi_i$ ):
\begin{eqnarray}
&&\chi_1' -1 =0,\ \ \ \ \chi_2' -i =0,\nonumber\\
&& \chi_3' -1 =0,\ \ \ \ \chi_4' +i =0,\nonumber\\
&&\chi_1 +i\chi_2 =0,\ \ \chi_3 -i\chi_4 =0.
\end{eqnarray}
The solutions of the above equations satisfying the boundary conditions given in
(\ref{ka}) are
\begin{eqnarray}
 \chi_1 =+ \kappa,\ \  \chi_2 =+i \kappa,\ \  \chi_3 = \kappa,\ \  \chi_4 = -i\kappa.
\end{eqnarray}
Now, with the help of these identifications of $\chi_i$, the
expression of $S_1$ given in (\ref{s1}) reduces to the following:
\begin{eqnarray}
S_1[\phi(x,\kappa), \kappa]  &=&\int d^4 x \left [ \kappa  B_\omega \cdot \partial^\mu A_\mu  +  i\kappa  \bar{ \mathbb 
C}_\omega \partial^\mu D_\mu [A]{ \mathbb C}_\omega +
 \kappa  B_\theta \cdot D^\mu[V]  X_\mu  \right.\nonumber\\
&- &\left. i\kappa  
\bar{ \mathbb C}_\theta \cdot D^\mu[ V-  X]D_\mu[ V+ X]  ({ \mathbb C}_\theta -{ 
\mathbb C}_\omega )
\right],
\end{eqnarray}
 which vanishes at $\kappa =0$. However at $\kappa =1$, it is nothing but the gauge-fixing and ghost parts of the action given in Eq. (\ref{action}),
 i.e.,
 \begin{eqnarray}
S_1 [\phi (x,1),1]&=&\int d^4 x \left [  B_\omega \cdot \partial^\mu A_\mu  +i \bar{ \mathbb 
C}_\omega \partial^\mu D_\mu [A]{ \mathbb C}_\omega +B_\theta \cdot D^\mu[V]  X_\mu \right.\nonumber\\
&-&\left. i  
\bar{ \mathbb C}_\theta \cdot D^\mu[ V-  X]D_\mu[ V+ X]  ({ \mathbb C}_\theta -{ 
\mathbb C}_\omega )\right], \nonumber\\
&=&\int d^4 x\ {\cal L}_{  GF+FP}.
 \end{eqnarray}
This shows that  the gauge-fixing and ghost terms appear naturally within functional integral through Jacobian calculation under FFBRST transformation with finite field-dependent 
parameter obtainable from (\ref{thet})  as
\begin{eqnarray}
\int D\phi \stackrel{FFBRST}{----\longrightarrow} \int D\phi\ e^{-S_1}=\int D\phi\ e^{-\int d^4 x\ {\cal L}_{  GF+FP}}.
\end{eqnarray}
Thus, we are able to say that the gauge-fixing and corresponding Faddeev-Popov 
ghost terms are nothing but the Jacobian of path integral measure of
partition function
under FFBRST transformation with appropriate parameter. 

\section{Concluding remarks}
In this Letter, we have considered the CFN decomposed $SU(2)$ YM theory
having different ghost structures,
which plays an important role in explaining low-energy limit of the theory,
and   analysed the BRST symmetry of this
theory. Further,
we have generalized the nilpotent BRST transformations of
CFN variables in Euclidean space. The generalization has been made by making the infinitesimal
Grassmannian parameter of BRST transformation  finite and field-dependent
which is known FFBRST transformation. We have shown that the
FFBRST transformations are symmetry 
of the effective action however these do not leave the
partition function (functional integral) invariant
because the path integral measure is not invariant under these.
 The Jacobian 
of path integral measure in the expression of functional 
 integral changes non-trivially
 under FFBRST transformations. Further, the method of Jacobian evaluation has been discussed. Utilizing this method, we have calculated the Jacobian
 for path integral measure under FFBRST transformations
 for a particular  choice of finite field-dependent parameter. It has been found  that Jacobian of path integral measure
of YM theory analysed in CFN variables   under FFBRST transformations  for this particular choice of  
 parameter 
 produces the gauge-fixing and faddeev-popov ghost terms in the effective theory. 
 However,   for a different choices 
 of finite field-dependent parameter the FFBRST transformation will
 lead to connection between 
 MAG and another Abelian gauge of the theory. 
 From my points of view, these results will hold for any general gauge theory, even though, we have shown it for a CFN decomposed
 YM theory. It will be interesting to generalize the results
 for very general  gauge theory, including those with open or reducible 
 gauge algebras.

\end{document}